\documentclass[12pt]{article}
\usepackage{amsmath}
\usepackage{bm}
\usepackage{amssymb}

\newcommand{\HM}{M}

\newcommand{\cN}{{\mathcal{N}}}

\newcommand{\mfah}{\hat{\mathfrak{h}}}

\newcommand{\FLS}{\mathfrak{L}\!\!\!\,\!\mathfrak{F}}

\def\z#1{{{\zeta_#1}}}
\def\h#1#2{{{{\mathrm h}_{#1#2}}}}

\begin{document}

\begin{center}
{\Large {\bf BFKL pomeron\\[3mm] in the next-to-next-to-leading approximation\\[4mm] in the planar N=4 SYM theory}}

\vspace{8mm}

{\sc
V.~N.~Velizhanin}\\[8mm]

{\it Theoretical Physics Division\\
NRC ``Kurchatov Institute''\\
Petersburg Nuclear Physics Institute\\
Orlova Roscha, Gatchina\\
188300 St.~Petersburg, Russia}\\[1mm]
and\\[1mm]
{\it Institut f{\"u}r  Mathematik und Institut f{\"u}r Physik\\
Humboldt-Universit{\"a}t zu Berlin\\
IRIS Adlershof, Zum Gro\ss{}en Windkanal 6\\
12489 Berlin, Germany
}\\[7mm]

\textbf{Abstract}\\[2mm]
\end{center}
\noindent{
We find the eigenvalue of the kernel of BFKL equation in the next-to-next-to-leading logarithm approximation in the planar $\cN=4$ SM theory from the constraints, coming from the six-loop anomalous dimension of twist-2 operators and known large-$\gamma$ limit.
}
\newpage

\setcounter{page}{1}

\section{Introduction}

The Balitsky-Fadin-Kuraev-Lipatov (BFKL) equation~\cite{Lipatov:1976zz,Kuraev:1977fs,Balitsky:1978ic,Fadin:1998py} was obtained during the study of the Regge processes at high energies $\sqrt{s}$ in the non-abelian gauge theories.
In this kinematics, when a transferring momentum is very small, the large logarithms appear and they should be summed in all orders of perturbative theory.
Thus, the BFKL equation in the leading-logarithm approximation sums all leading logarithmic terms $(\alpha_s\ln 1/x)^\ell$ in all order of the perturbative theory. In this leading approximation only ladder diagrams give the contribution, which can be evaluated with the Sudakov decomposition order by order and it is possible to write such corrections in a general form with the help of Bethe-Salpeter equation for the partial wave, which is known as BFKL equation.
As described, for example, in Ref.~\cite{Fadin:1998py}, the BFKL equation allows to estimate the total cross-section $\sigma (s)$ for the high energy scattering of colourless particles $A,B$ 
\begin{equation}
\sigma (s)=
\int \frac{d^2q_1}{2\pi q_1^2}
\int \frac{d^2q_2}{2\pi q_2^2}
\,\Phi _A(\vec{q}_1)
\,\,\Phi _B(\vec{q}_2)
\,\int_{a-i\infty }^{a+i\infty }\frac{d\omega }{2\pi i}
\,\left(
\frac{s}{q_1\,q_2}
\right) ^\omega \,G_\omega (\vec{q}_1,\vec{q}_2)\,,  \label{CrossS}
\end{equation}
where $G_\omega (\vec{q}_1,\vec{q}_2)$ is the $t$-channel partial wave for the reggeized gluon scattering at $t=0$ and $\vec{q}_1$ and $\vec{q}_2$ are transverse momenta of gluons with the virtualities $-\vec{q}_1^{\;2}\equiv -q_1^{2}$ and $-\vec{q}_2^{\;2}\equiv -q_2^2$ correspondingly, $s=2p_Ap_B$ is the squared invariant mass of the colliding particles with momenta $p_A$ and $p_B$. 
The generalized BFKL equation for $G_\omega (\vec{q}_1,\vec{q}_2)$ in the following form
\begin{equation}
\omega \,G_\omega (\vec{q}_1,\vec{q}_2)=\delta
^{D-2}(\vec{q}_1-\vec{q}_2)+\int d^{D-2}
{q}\ K(\vec{q}_1,\vec{{q}})\,G_\omega (\vec{{q}},\vec{q}_2)\,.
\label{l2}
\end{equation}
Here 
\begin{equation}
K(\vec{q}_1,\vec{{q}}_2)=2\,\omega (q_1)\,\delta^{(D-2)}(\vec{q}_1-\vec{q}_2)
+K_r(\vec{q}_1,\vec{q}_2)\,.  \label{l3}
\end{equation}
The gluon Regge trajectory $\omega (q)$ and the integral kernel $K_r(\vec{q}_1,\vec{q}_2)$ are expanded in the series over the QCD coupling constant 
\begin{equation}
\omega (q)=\omega _B(q)+\omega ^{(2)}(q)+...\,,\qquad 
K_r(\vec{q}_1,\vec{q}_2)=K_r^B(\vec{q}_1,\vec{q}_2)
+K_r^{(1)}(\vec{q}_1,\vec{q}_2)+...\,\,,\,.
\label{l4}
\end{equation}
The gluon Regge trajectory $\omega(q)$) and the integral kernel $K_r^{(1)}(\vec{q}_1,\vec{q}_2)$ can be found up to the next-to-leading logarithm approximation in Ref.~\cite{Fadin:1998py}.

As it was shown in~\cite{Balitsky:1978ic}, a complete and orthogonal set of eigenfunctions of the homogeneous BFKL equation in LLA is 
\begin{equation}
G_{n,\gamma }(\vec{q},\vec{q}_2)\ =\ 
\left( \frac{q^{2}}{q_2^{2}}\right) ^{\gamma -1}
\end{equation}

The BFKL kernel in this representation is diagonalized up to the effects
related with the running coupling constant $\alpha _{s}(q^{2})$: 
\begin{equation}
\omega =\frac{\alpha _{s}(q^{2})N_{c}}{\pi }\biggl[ \chi (n,\gamma )+\delta
(n,\gamma )\frac{\alpha _{s}(q^{2})N
_{c}}{4\pi }\biggr] \,.
\end{equation}

To find the eigenvalue of the kernel of BFKL equation one can used the eigenfunctions $q_1^{2(\gamma -1)}$ of the Born kernel:
\begin{equation}
\omega=\int d^{D-2}q\ K(\vec{q}_1,\vec{q})
\left(\frac{q^2}{q_1^2}\right) ^{\gamma-1}=\frac{\alpha _s(q_1^2)\,N_c\,}\pi
\left( \chi (\gamma )+\delta (\gamma )\frac{\alpha _s(q_1^2)N_c}{4\,\pi }
\right) \,,\, \label{l12}
\end{equation}
The quantity $\chi (\gamma ) $ is proportional to the eigenvalue of the Born kernel
\begin{equation}
 \chi (\gamma )=2\Psi (1)-\Psi (\gamma )-\Psi (1-\gamma )\,,\,\,\,\Psi
 (\gamma )=\Gamma ^{\prime }(\gamma )/\Gamma (\gamma )\,,  \label{l13}
\end{equation}
 and  the correction $\delta (\gamma)$ is given by~\cite{Fadin:1998py}
\begin{eqnarray}
\delta (\gamma )&=&-\left[ \left( \frac{11}3-\frac{2n_f}{3N_c}\right) \frac
12\left( \chi ^2(\gamma )-\Psi ^{\prime }(\gamma )+\Psi ^{\prime }(1-\gamma
)\right) -\left( \frac{67}9-\frac{\pi ^2}3-\frac{10}9\frac{n_f}{N_c}\right)
\chi (\gamma )\right. 
\nonumber\\[2mm]&&
\left. -6\zeta (3)+\frac{\pi ^2\cos(\pi \gamma )}{\sin^2(\pi \gamma
)(1-2\gamma )}\left( 3+\left( 1+\frac{n_f}{N_c^3}\right) \frac{2+3\gamma
(1-\gamma )}{(3-2\gamma )(1+2\gamma )}\right) \right. 
\nonumber\\[2mm]&&
\left. -\Psi ^{\prime \prime }(\gamma )-\Psi ^{\prime \prime }(1-\gamma )
-
\frac{\pi ^3}{\sin (\pi \gamma )}+4\phi (\gamma )\right] \,.  \label{l12a}
\end{eqnarray}
The function $\phi(\gamma )$ is 
\begin{eqnarray}
\phi (\gamma )&=&-\int_0^1\frac{dx}{1+x}\left( x^{\gamma -1}+x^{-\gamma
}\right) \int_x^1\frac{dt}t\ln (1-t) \nonumber\\[2mm]
&=&\sum_{n=0}^\infty (-1)^n\left[ \frac{\Psi (n+1+\gamma )-\Psi (1)}{(n+\gamma
)^2}+\frac{\Psi (n+2-\gamma )-\Psi (1)}{(n+1-\gamma )^2}\right] \,.
\label{l14}
\end{eqnarray}

The BFKL equation in the leading logarithm approximation is the same in any gauge theory and has a lot of remarkable properties.
For example, the integrability in the quantum field theory was firstly discovered by L.N. Lipatov during the study of the BFKL equation~\cite{Lipatov:1993yb,Lipatov:1994xy}. 

The generalisation of the computations of the BFKL equation in the next-to-leading-logarithm approximation, performed by  V.S. Fadin and L.N. Lipatov in QCD~\cite{Fadin:1998py}, to the maximally extended $\cN=4$ supersymetric Yang-Mills (SYM) theory shows~\cite{Kotikov:2000pm}, that a lot of terms in the QCD result~(\ref{l12a}) are cancelled and the final result contains the functions, which have the same property called later as a transcedentality\footnote{Their large $\gamma$ limit coincides with the special transcendental numbers such as zeta-numbers $\zeta_i$}:
\begin{eqnarray}
\delta (\gamma )&=&
\Psi ^{\prime \prime }(\gamma )+\Psi ^{\prime \prime }(1-\gamma )
+6\zeta (3)
+
\frac{\pi ^3}{\sin (\pi \gamma )}
-4\phi (\gamma ) \,.  
\label{BFKLNLLA}
\end{eqnarray}
Using the same suggestion the results for the anomalous dimension of the twist-2 operators in $\cN=4$ SYM theory was obtained without any computation~\cite{Kotikov:2002ab}, but argued from the relation between BFKL and Dokshitzer-Gribov-Lipatov-Altarelli-Parizi (DGLAP)~\cite{Gribov:1972ri,Altarelli:1977zs,Dokshitzer:1977sg} equations. The maximal transcedentality principle was confirmed by the direct diagrammatic calculations at two loops~\cite{Kotikov:2003fb} and then successfully used for the finding the three-loop anomalous dimension~\cite{Kotikov:2004er} from the corresponding result, computed directly in QCD~\cite{Moch:2004pa}.
This result help to confirm a general form of the asymptotic Bethe-ansatz~\cite{Beisert:2004hm}, which can be used for the computations of the anomalous dimension of composite operators in the $\cN=4$ SYM theory. Then this maximal transcedentality principle was used for the computations of the general form of the anomalous dimension for twist-2 operators as with the help of integrability~\cite{Staudacher:2004tk,Kotikov:2007cy,Bajnok:2008qj,Lukowski:2009ce,Marboe:2014sya} as from the constraints coming from the generalised double-logarithmic equation~\cite{Velizhanin:2011pb,Velizhanin:2013vla}. It is no doubt, that the maximal transcedentality principle works for the BFKL equation at higher orders too, but the direct diagrammatic computations were very cumbersome, while a number of available constraints was not enough to perform a reconstruction procedure similar what we used for the anomalous dimension. However, sometimes ago we obtain the general result for the six-loop anomalous dimension of the twist-2 operators in the planar $\cN=4$ SYM theory~\cite{Marboe:2014sya} in the collaboration with C. Marboe and D. Volin using their powerful code~\cite{Marboe:2014gma}, which realised a recently proposed method for the computations of the anomalous dimension of composite operators~\cite{Gromov:2013pga}. Being analytically continued the six-loop anomalous dimension provide us with a lot of information about BFKL equation, so, we can try to use this new information for the computations of the eigenvalue of the kernel of BFKL equation in the next-to-next-to-leading approximation (NNLLA).

\section{Expansion of anomalous dimension near $M=-1+\omega$}

The eigenvalue of the kernel of BFKL equation relates the small correction to the BFKL-pomeron with the anomalous dimension of the twist-2 operators, which is perturbative corrections to the canonical dimension of the composite operators. The anomalous dimension should be evaluated near this point $j=1$, where the anomalous dimension for the gluon operator in QCD has the poles. In the $\cN=4$ SYM theory $M=j-2$, so, to make a link between the BFKL equation and DGLAP equation we should evaluate the known six-loop anomalous dimension near $M=-1+\omega$, where $\omega$ is exactly the same parameter as in the left-hand side of the BFKL equation~(\ref{l12}). 

Expanding BFKL equation~(\ref{l12}) order by order in $\gamma$ and substitute anomalous dimension into the right hand side we indeed find that all highest poles are cancelled. In this way one can obtain predictions for the highest poles for the anomalous dimension in any order of perturbative theory. That is, the BFKL equation fixes all coefficients in the following expansion of the anomalous dimension near $M=-1+\omega$:
\begin{eqnarray}
\gamma(M)\big|_{M=-1+\omega}^{\mathrm{LLA}}&=&
\sum_{\ell=1}c^{\mathrm{LLA}}_{\ell}\left(\frac{g^2}{\omega}\right)^\ell+\ \ldots\,,
\\&=&
-\ 2\left(\frac{4g^2}{\omega}\right)
+0\left(\frac{4g^2}{\omega}\right)^2
+0\left(\frac{4g^2}{\omega}\right)^3
-4 \z3\left(\frac{4g^2}{\omega}\right)^4
+0\left(\frac{4g^2}{\omega}\right)^5\nonumber\\
&& 
-\ 4 \z5\left(\frac{4g^2}{\omega}\right)^6
- 24 \z3^2\left(\frac{4g^2}{\omega}\right)^7
-4 \z7\left(\frac{4g^2}{\omega}\right)^8
-64 \z3 \z5\left(\frac{4g^2}{\omega}\right)^9
\end{eqnarray}
where coefficient $c^{\mathrm{LLA}}_{\ell}$ has the transcendentality equal to $\ell-1$.
In the next-to-leading logarithms approximation (NLLA) the BFKL equation deal with the subleading logarithms and fix all coefficients in the following expansion:
\begin{equation}
\gamma(M)\big|^{\mathrm{NLLA}}_{M=-1+\omega}=\sum_{\ell=1}c^{\mathrm{NLLA}}_{\ell}\left(\frac{g^2}{\omega}\right)^\ell\!\omega\ +\ \ldots
\end{equation}
and in general the ${{\mathrm{N}}{}^k{\mathrm{LLA}}}$ BFKL equation gives
\begin{equation}
\gamma(M)\big|^{{\mathrm{N}}{}^k{\mathrm{LLA}}}_{M=-1+\omega}=\sum_{\ell=1}c^{{\mathrm{N}}{}^k{\mathrm{LLA}}}_{\ell}\left(\frac{g^2}{\omega}\right)^\ell\omega^k\ +\ \ldots
\end{equation}

From the BFKL equation we know all highest poles of the anomalous dimension in any order of perturbative theory, but from the available up to now the NLLA result we know only two highest. At the same time we know expansion over $\omega$ for the six-loop anomalous dimension for the first twelve coefficients in the expansion\footnote{This restriction is related with the large $j$ limit for the harmonic sums, which is available now up to level~12 from Ref.~\cite{Blumlein:2009cf}, i.e. for $\zeta_{12}$ and similar special numbers.}. That is we know information about ${{\mathrm{N}}{}^k{\mathrm{LLA}}}$ BFKL equation up to $k=11$. Than, we can use such information for the reconstruction of the BFKL equation in ${{\mathrm{N}}{}^k{\mathrm{LLA}}}$.  In this paper we will use such information for the reconstruction of the eigenvalue of the kernel BFKL equation in the next-to-next-to-leading approximation (${{\mathrm{N}}{}^2{\mathrm{LLA}}}$). The expansion of the six-loop anomalous dimension up to necessary order has the following form:
\begin{eqnarray}
\gamma(M)\big|_{M=-1+\omega}&=&\left(
-2
+2\, {\z2}\, \omega^2
\right)\left(\frac{4g^2}{\omega}\right)
+4 \, {\z3}\, \omega^2\left(\frac{4g^2}{\omega}\right)^2
+ \left(
-{\z3}\,{\omega}
+\frac{29}{4}\,{\z4}\,\omega^2
\right)\left(\frac{4g^2}{\omega}\right)^3
\nonumber\\&&
+\left(
-4\, {\z3}
-\frac{5}{4}\, {\z4}\,\omega
+\left(5\, {\z2}\, {\z3}+\frac{77}{4}\, {\z5}\right)\omega^2
\right)\left(\frac{4g^2}{\omega}\right)^4
\nonumber\\&&
+\left(
\Big(2\, {\z2}\, {\z3}
+16\, {\z5}\Big)\omega
+\left(21\,{\z3}^2
+\frac{61}{3}\, {\z6}\right)\omega^2\right)
\left(\frac{4g^2}{\omega}\right)^5
\nonumber\\&&\hspace*{-25mm}
+ \left(
-4\, {\z5}
+\left(\frac{143}{48} {\z6}
-3 {\z3}^2\right){\omega}
+\left(8\, {\z2}\, {\z5}
+\frac{277}{8}\, {\z3}\, {\z4}
-\frac{631}{32}\, {\z7}\right){\omega^2}
\right)\left(\frac{4g^2}{\omega}\right)^6\,.
\label{6loopsomega}
\end{eqnarray}

\section{Small $\gamma$ expansion from constraints}

From Eq.~(\ref{6loopsomega}) we can easily obtain the expansion of $\omega$ over $\gamma$ up to third order of perturbative theory.
Remind, that the anomalous dimension is related with $\omega$ in the so-called non-symmetric point, that is, for the $\gamma$ shifted on $\omega$: $\gamma\to\gamma+\omega$ (see, for example, Ref.~\cite{Fadin:1998py} for details). This expansion demands a resubstitution of $\omega$ in the right hand side of Eq.~(\ref{l12}) and a reexpansion over the coupling constant up to necessary order. This procedure looks like a reciprocity procedure for the anomalous dimension~\cite{Dokshitzer:2005bf,Dokshitzer:2006nm}, where the argument of the harmonic sums~$M$, which enter into the expression for $\gamma$, are shifted on this $\gamma$ ($M=M+\gamma$). Major formal difference of such reexpansions is that for the anomalous dimension the reciprocity is some additional procedure to the original computations (which can be done, for example, with the help of ABA or QSC approach), while for the BFKL equation such procedure should be taken into account before comparison with the anomalous dimension, if one can obtain some result separately for the holomorphic and antiholomorphic parts.
So, if we write for the $\omega$ the following general expansion over $\gamma$
\begin{eqnarray}
\omega=\sum_{\ell=1}
\left(
{\mathfrak{F}\!\!\mathfrak{L}}^{(\ell)}\!\left(-\frac{\gamma}{2}\right) 
+
{\mathfrak{F}\!\!\mathfrak{L}}^{(\ell)}\!\left(1+\frac{\gamma}{2}\right) 
\right)
g^{2\ell}
\end{eqnarray}
we should perform the substitution $\gamma\to\gamma+\omega$ and then expanding the following expression
\begin{eqnarray}
\omega&=&
g^{2}\left(
{\mathfrak{F}\!\!\mathfrak{L}}^{(0)}\!\left(-\frac{\gamma+\omega}{2}\right) 
+
{\mathfrak{F}\!\!\mathfrak{L}}^{(0)}\!\left(1+\frac{\gamma+\omega}{2}\right) 
\right)
\nonumber\\
&&+\ g^{4}\left(
{\mathfrak{F}\!\!\mathfrak{L}}^{(1)}\!\left(-\frac{\gamma+\omega}{2}\right) 
+
{\mathfrak{F}\!\!\mathfrak{L}}^{(1)}\!\left(1+\frac{\gamma+\omega}{2}\right) 
\right)
+g^{6}
\sum_{k=-3}
\widehat{\mathfrak{F}\!\!\mathfrak{L}}_k^{(2)}\,\gamma^k
\label{IptuptoNNLLA}\\
&=&
g^{2}\left(
{\mathfrak{F}\!\!\mathfrak{L}}^{(0)}\!\left(-\frac{\gamma}{2}\right) 
+
{\mathfrak{F}\!\!\mathfrak{L}}^{(0)}\!\left(1+\frac{\gamma}{2}\right) 
\right)
\nonumber\\
&&+\ g^{4}\left(
{\mathfrak{F}\!\!\mathfrak{L}}^{(1)}\!\left(-\frac{\gamma}{2}\right) 
+
{\mathfrak{F}\!\!\mathfrak{L}}^{(1)}\!\left(1+\frac{\gamma}{2}\right) 
\right)
+g^{6}
\sum_{k=-5}
{\mathfrak{F}\!\!\mathfrak{L}}_k^{(2)}\,\gamma^k
\label{IptuptoNNLLA2}
\end{eqnarray}
up to third order of the perturbative theory we can find coefficients ${\mathfrak{F}\!\!\mathfrak{L}}_k^{(2)}$
of the expansion of the NNLLA corrections to the eigenvalue of the BFKL-pomeron
\begin{equation}
\sum_{k=-5}^{\infty}
{\mathfrak{F}\!\!\mathfrak{L}}_k^{(2)}\,\gamma^k
=
{\mathfrak{F}\!\!\mathfrak{L}}^{(2)}\!\left(-\frac{\gamma}{2}\right) 
+
{\mathfrak{F}\!\!\mathfrak{L}}^{(2)}\!\left(1+\frac{\gamma}{2}\right) \,.
\label{IptNNLLA}
\end{equation}
Note, that the difference between the coefficients $\widehat{\mathfrak{F}\!\!\mathfrak{L}}_k^{(2)}$ and ${\mathfrak{F}\!\!\mathfrak{L}}_k^{(2)}$ comes from the shifting in the argument $\gamma\to\gamma+\omega$ (from the first two terms in Eq.~(\ref{IptuptoNNLLA})) and the expansion of $\omega$ in Eq.~(\ref{IptuptoNNLLA}) has maximally only single logarithms (i.e. the negative powers of $\gamma$ no more than $(g^2/\gamma)^\ell$), while Eq.~(\ref{IptNNLLA}) has the double logarithms (i.e. $\gamma\,(g^2/\gamma^2)^\ell$).

Substitute Eq.~(\ref{6loopsomega}) into Eq.~(\ref{IptuptoNNLLA}) we find the following expansion of the $\omega$ in third order of the perturbative expansion:
\begin{eqnarray}
\omega&=&
+g^2 
\left(
\frac{8}{\gamma }
-2\, {\z3}\, \gamma ^2
-\frac{{\z5}}{2} \gamma ^4
-\frac{{\z7}}{8} \gamma ^6
\right)\nonumber\\&&
+g^4 
\bigg(
-\frac{64}{\gamma ^3}
-24\, {\z3}
+5\, {\z4}\, \gamma
+\gamma ^2 
\Big(
4\, {\z2}\,{\z3}
+20\, {\z5}
\Big)
+\gamma ^3 
\left(
3\, {\z3}^2
-\frac{143}{48} {\z6}
\right)\nonumber\\&&
\qquad+\gamma ^4 
\Big(
{\z2}\, {\z5}
+14\, {\z7}
\Big)
\bigg)\nonumber\\&&
+g^6 \bigg(
\frac{1024}{\gamma ^5}
-\frac{512 }{\gamma ^3}{\z2}
+\frac{576}{\gamma ^2} {\z3}
-\frac{464}{\gamma } {\z4}
+840 {\z5}
+64 {\z2} {\z3}
+\gamma  
\Big(
-40 {\z3}^2
-373 {\z6}
\Big)\nonumber\\&&
\qquad+\gamma ^2 
\left(
-8 {\z2} {\z5}
-86 {\z3} {\z4}
+\frac{1001}{4} {\z7}
\right)
\bigg)\,.
\label{OmegaT7}
\end{eqnarray}

\subsection{Large $\gamma$ limit}

Studying the properties of the eigenvalue of the kernel of BFKL equation we have found, that its large $\gamma$ limit coincide with the large $\HM$ limit for the anomalous dimension up to the finite part. We thinking about the possibility to use this information for the reconstruction, but it was not clear, whether such relation should work or not in the next orders. However recently the general formulae for the large $\gamma$ limit was obtained~\cite{Basso:2014pla}\footnote{I thank L.N. Lipatov for the information about this result.} and indeed the expressions, which can be obtained from this equation and the large $\HM$ limit for the anomalous dimension coincide.
Thus, the information about large $\gamma$ limit includes also in the list of our constraints.

\section{Construction of the basis}

The main problem for the reconstruction procedure, which we are going to use, is a construction of the optimal (minimal) basis. To construct such basis we have information from two sources: the BFKL equation in the next-to-leading logarithm approximation and the anomalous dimension of twist-2 operators. From the study of the anomalous dimension we know, that only the usual harmonic sums, defined as~\cite{Vermaseren:1998uu}:
\begin{equation}
S_{a,b,c,\ldots}(M)
=\sum_{i=1}^M \frac{({\rm sgn}(a))^{i}}{i^{|a|}} S_{b,c,\ldots}(i),
\qquad\qquad
S_{a}(M)
=\sum_{i=1}^M \frac{({\rm sgn}(a))^{i}}{i^{|a|}}
\label{HS}
\end{equation}
 can enter into the final expression.
So, following a suggestion about the deep relation between the BFKL and DGLAP equations~\cite{Kotikov:2002ab} we suggest, that the similar sums, or, more precisely, their analytical continuations, will enter into the basis for the eigenvalues of the BFKL kernel in the higher order corrections. However, looking on the available result for the BFKL equation~(\ref{l12a}) one can find some unusual term, which can be rewritten through the analytically continued harmonic sum $S_{-1}$, 
\begin{equation}
\frac{\pi^2}{\sin\big(\pi\gamma/2)}=\beta\left(-\frac{\gamma}{2}\right)+\beta\left(1+\frac{\gamma}{2}\right),
\qquad
S_{-1}\rightarrow^{{}^{\hspace*{-5mm}{\mathrm{A.C.}}}}
\beta(z)=
\Psi\bigg(\frac{z-1}{2}\bigg)
-
\Psi\bigg(\frac{z}{2}\bigg)
\label{beta0}
\end{equation}
Remind, that the results for the anomalous dimension of twist-2 operators do not contain such sums (with index $(-1)$), what allows considerably reduce the basis. For the BFKL equation such sum appears and in spite of it is multiplied to $\z2$, unfortunately, it is not clear how to generalise this procedure to the higher order corrections. However, there is another representation of the same result for the NLLA corrections, which is differed only in the last terms due to the following equality~(see Ref.~\cite{Kotikov:2000pm} for details):
\begin{eqnarray}
&&
\frac{\pi ^3}{\sin (\pi \gamma )}+4\phi (\gamma )=
-\int_0^1\frac{dx}{1+x}\left( x^{\gamma -1}+x^{-\gamma
}\right) \int_x^1\frac{dt}t\ln (1-t) 
\\
&&
=\frac{\pi ^3}{\sin (\pi \gamma )}+4\sum_{n=0}^\infty (-1)^n\left[ \frac{\psi (n+1+\gamma )-\psi (1)}{(n+\gamma
)^2}+\frac{\psi (n+2-\gamma )-\psi (1)}{(n+1-\gamma )^2}\right] 
\\
&&
=\ 
2\sum_{k=0}^{\infty }\frac{(-1)^{k}}{k+\gamma}(-1)^{k}\beta ^{\prime }(k+1)
+2\sum_{k=0}^{\infty }\frac{(-1)^{k}}{k-\gamma+1}(-1)^{k}\beta ^{\prime }(k+1),
\label{HSbeta}
\end{eqnarray}
where
\begin{equation}
\beta ^{\prime }(z)=\frac{1}{4}\Biggl[ \Psi ^{\prime }\Bigl(\frac{z+1}{2}%
\Bigr)-\Psi ^{\prime }\Bigl(\frac{z}{2}\Bigr)\Biggr]
\label{beta}
\end{equation}
Owing to this simple transformation for the last terms two simplifications appear: the unusual term disappears and in the obtained function $\gamma$ enters only into the first denominator. 
As we expect, that integrand can be represented as some polynomial over $x$ such property will hold true in the higher orders too. Thus, instead of having to work with the usual harmonic sums, which should be then analytically continued for the expansion over $\gamma$, it is much simpler to work directly with such functions, which expansion over $\gamma$ is rather trivial and can be found in Appendix.
Note also, that $\beta'(z)$ from Eq.~(\ref{beta}) is related with the analytical continuation of the harmonic sum $S_{-2}$ through (see Ref.~\cite{Kotikov:2005gr} for details)
\begin{equation}
S_{-2}(M)= S_{-2}(\infty)-(-1)^M\beta'(M+1),\qquad 
S_{-2}(\infty)=\zeta_{-2}=\left(\frac{1}{2^1}-1\right)\z2=-\frac{\z2}{2}
\label{HSbetar}
\end{equation}
and if we substitute the expression for the $\beta'(M+1)$ from Eq.~(\ref{HSbetar}) into Eq.~(\ref{HSbeta}) we can rewrite the sum in the following form
\begin{eqnarray}
\sum_{k=0}^{\infty }\frac{(-1)^{k}}{k+\gamma}(-1)^{k}\beta ^{\prime }(k+1)&=&
-\sum_{k=0}^{\infty }\frac{(-1)^{k}}{k+\gamma}S_{-2}(k)
+\sum_{k=0}^{\infty }\frac{(-1)^{k}}{k+\gamma}\left(-\frac{\z2}{2}\right)\nonumber\\
&=&-\FLS_{-1,-2}(\gamma)-\frac{\z2}{2}\FLS_{-1}(\gamma)
\end{eqnarray}
and due to such relation we will denote sums in the basis with the first index $(-1)$ (for example, the sum in Eq.~(\ref{HSbeta}) we will denote as $\FLS_{-1,-2}$). 

Keep in mind the maximal transcendentality principle~\cite{Kotikov:2002ab} we will interesting for the construction of the basis for the NNLLA corrections only with the functions which have the transcendentality $5$. The most complicated sum, which is an analogue of the usual harmonic sum $S_{-2,1,1,1}$ appearing in the expression for the three-loop anomalous dimension of twist-2 operators, can be constructed in the following way:
\begin{eqnarray}
\sum_{k=0}^{\infty }\frac{(-1)^{k}}{k+\gamma}S_{-2,1,1}(k)
&=&
\sum_{k=0}^{\infty }\frac{(-1)^{k}}{k+\gamma}
\sum_{n=1}^{k}\frac{(-1)^{n}}{n}S_{1,1}(n)
\overset{{}^{2S_{1,1}=S_2+S_1^2}}{\simeq}\nonumber\\
&\simeq&
\sum_{k=0}^{\infty }\frac{(-1)^{k}}{k+\gamma}
\sum_{n=1}^{k}\frac{(-1)^{n}}{n}S_{1}(n)^2
+
\sum_{k=0}^{\infty }\frac{(-1)^{k}}{k+\gamma}
\sum_{n=1}^{k}\frac{(-1)^{n}}{n}S_{2}(n)\qquad \\
&\Rightarrow &
\sum_{k=0}^{\infty }\frac{(-1)^{k}}{k+\gamma}(-1)^{k}\beta ^{\prime }(k+1)
(\Psi(k+1)-\Psi(1))^2\nonumber\\&&+
\sum_{k=0}^{\infty }\frac{(-1)^{k}}{k+\gamma}(-1)^{k}\beta ^{\prime }(k+1)
(\Psi'(k+1)-\Psi'(1))\\&=&
\FLS_{-1,-2,{1,1}}(\gamma)+
\FLS_{-1,-2,2}(\gamma),
\end{eqnarray} 
where we used the relation between harmonic sums $2S_{1,1}=S_2+S_1^2$ to avoid an additional summation.
The final most general basis for the eigenvalue of the kernel of BFKL equation in the next-to-next-to-leading approximation along with the expansion of all functions up to necessary order can be found in Appendix.

\subsection{Reconstruction}

The reconstruction is performed in the same way as in our previous papers~\cite{Velizhanin:2010cm,Velizhanin:2011pb,Velizhanin:2012nm,Velizhanin:2013vla,Marboe:2014sya}. Namely, we believe, that the coefficients in ansatz, which can be constructed from the functions, discussed in the previous section, should be the integer numbers. This means, that each equation in the obtained system will be a Diophantine equation. This system of Diophantine equations has a practical interest (for example, a knapsack problem) and can be solved with the method from a number theory even when the rank of the system is less (considerable) than its dimension. The most powerful method is related with the usage of a \texttt{LLL}-algorithm~\cite{Lenstra:1982}, which being applied to a matrix gives the new matrix in which all rows arranged with a minimal Euclidian norm.

From the six-loops anomalous dimension and large $\gamma$ limit we have $19$ constraints, while the general basis contains $63$ functions. Moreover, the coefficients in the obtained equations are rather small numbers, then the \texttt{LLL}-algorithm will not work for such large basis. However, trying to reduce the basis we reanalyse the expression for the three-loops anomalous dimension and suggest, that the minimal basis in our case should contain for the product of the several functions only one function with the negative index. In this way our basis is reduced to the following $29$ functions:
\begin{eqnarray}
{\mathrm{Basis}}\,\Big|_{\mathrm{NNLLA}}&=&
\Big\{\FLS_{-1,-2,1,1},
\FLS_{-1,1,-2,1},
\FLS_{-1,-2,2},
\FLS_{-1,2,-2},
\FLS_{-1,-3,1},
\FLS_{-1,1,-3},
\FLS_{-1,-4},
\nonumber\\[2mm]&&\hspace*{-29mm}
\quad
{\FLS_{1}\otimes\FLS_{-1,-2,1}},
{\FLS_{1}\otimes\FLS_{-1,-3}},
{\FLS_{2}\otimes\FLS_{-1,-2}},
{\FLS_{1}\otimes\FLS_{-2}\otimes\FLS_{2}},
{\FLS_{1}\otimes\FLS_{-4}},
{\FLS_{2}\otimes\FLS_{-3}},
\nonumber\\[2mm]&&\hspace*{-29mm}
\quad
{\FLS_{-2}\otimes\FLS_{3}},
{\FLS_{-5}},
{\FLS_{5}},
{\z2\times\FLS_{-3}},
{\z2\times\FLS_{3}},
{\z2\times\FLS_{1}\otimes\FLS_{-2}},
{\z2\times\FLS_{-1}\otimes\FLS_{2}},
{\z2\times\FLS_{-1,-2}},
\nonumber\\[2mm]&&\hspace*{-29mm}
\quad
{\z2\times\FLS_{-1,2}},
{\z3\times\FLS_{-2}},
{\z3\times\FLS_{1}\otimes\FLS_{-1}},
{\z3\times\FLS_{-1,-1}},
{\z4\times\FLS_{-1}},
{\z4\times\FLS_{1}},
\z2\z3,
\z5\Big\}\,.
\end{eqnarray}
The coefficients for the last three functions in this basis can be fixed uniquely from the large $\gamma$ limit as all other functions do not contain such terms in this limit. 
From the obtained system of Diophantine equations, in which we multiply all functions with~$k$ indices by $(k-1)!$ and all functions with one index $a$ by $2^{|a|}$, we eliminate all variables related with the $\z4$ and $\z3$, as we suggest, that their coefficients may be the large numbers. 
We left with $11$ equation on $22$ variables.
Then, we construct a matrix from our system in the same way, as described in details in our previous paper~\cite{Velizhanin:2013vla}. 
Applying to the final matrix \texttt{LatticeReduce} function from \texttt{MATHEMATICA}, which realised \texttt{LLL}-algorithm, we have obtained the new matrix, in which only one row is the solution of the original non-uniform system of Diophantine equations:
\begin{equation}
{\mathrm{LLL}}=\{-6, 2, 1, -2, 10, -2, -12, 1, -2, 2, -2, -2, -1, 0, -1, 1, 16, 0, -8, 24, 4, 6\}
\label{LLLsolve}
\end{equation}
Missing coefficients for the functions, which we eliminate for the applicability of \texttt{LLL}-algorithm, can be easily obtained from the solution of the full original system with our solution~(\ref{LLLsolve}). 

Our final result for the eigenvalue of the kernel of BFKL equation in the next-to-next-to-leading logarithm approximation in the planar $\cN=4$ SYM theory can be written as:
\begin{eqnarray}
\left(
{\mathfrak{F}\!\!\mathfrak{L}}^{(2)}\!\left(-\frac{\gamma}{2}\right) 
+
{\mathfrak{F}\!\!\mathfrak{L}}^{(2)}\!\left(1+\frac{\gamma}{2}\right) 
\right)
&=&
6 {\z2} {\FLS}_{1,-2}
+6 {\z2} {\FLS}_{1,2}
-\frac{7}{2} {\z3} {\FLS}_{1,-1}
-12 {\FLS}_{1,-4}
\nonumber\\[2.29mm]
&&\hspace*{-50mm}
+20 {\FLS}_{1,-3,1}
-2 {\FLS}_{1,-2,2}
-4 {\FLS}_{1,1,-3}
-4 {\FLS}_{1,2,-2}
-36 {\FLS}_{1,-2,1,1}
+12 {\FLS}_{1,1,-2,1}
\nonumber\\[2.29mm]
&&\hspace*{-50mm}
+128 {\z2} {\FLS}_{-3}
+4 {\FLS}_1\otimes {\FLS}_{1,-2,1}
-4 {\FLS}_1\otimes {\FLS}_{1,-3}
+8 {\FLS}_2\otimes {\FLS}_{1,-2}
-64 {\z2} {\FLS}_{-2}\otimes {\FLS}_1 
\nonumber\\[2.29mm]
&&\hspace*{-50mm}
+192 {\z2} {\FLS}_{-1}\otimes {\FLS}_2
-64 {\FLS}_{-4}\otimes {\FLS}_1
-32 {\FLS}_{-3}\otimes {\FLS}_2
-64 {\FLS}_{-2}\otimes {\FLS}_1\otimes {\FLS}_2
\nonumber\\[2.29mm]
&&\hspace*{-50mm}
+32 {\FLS}_5
-32 {\FLS}_{-5}
+1080 {\z4} {\FLS}_{-1}
+88  {\z4} {\FLS}_1
+32 {\z2} {\z3}
+160 {\z5}\,.
\label{BFKLNNLLA}
\end{eqnarray}

The $\gamma^3$ term of the expansion of Eq.~(\ref{BFKLNNLLA})
\begin{equation}
\bigg(
\frac{1664 }{171}{\h53}
-\frac{8320}{57} {\h71}
+4 {\z2} {\z3}^2
-\frac{1542}{19} {\z3} {\z5}
-\frac{199373}{1368} {\z8}
\bigg)\,\gamma^3
\end{equation}
can be used for the prediction of the highest three poles of the seven-loop anomalous dimension near $M=-1+\omega$. For this purpose one should expand Eq.~(\ref{OmegaT7}) up to the transcendentality $8$ take into account Eq.~(\ref{IptuptoNNLLA}) and Eq.~(\ref{IptuptoNNLLA2}). 


\section{Conclusion and discussion}

In this paper we reconstruct the eigenvalue of the kernel of BFKL equation in the next-to-next-to-leading logarithms approximation (NNLLA) from the constraints, coming from the expansion of the six-loop anomalous dimension near $M=-1+\omega$ and large $\gamma$ limit solving the obtained system of the Diophantine equations with the help \texttt{LLL}-algorithm~\cite{Lenstra:1982}. 

Recently, the paper about NNLO corrections to the BFKL pomeron eigenvalue appeared~\cite{Gromov:2015vua}. In those paper authors compute with the help of QSC approach~\cite{Alfimov:2014bwa} some quantity, which they interpret as the eigenvalue of the BFKL pomeron. Unfortunately, authors did not provide any information how to use their result, so we can not compare its with the result obtained in the present paper. The QCS approach provides us with the very powerful method for the computations of the anomalous dimension and if the result of Ref.~\cite{Gromov:2015vua} indeed correct, that this gives the most simple way for the computations of the eigenvalue of the kernel of BFKL equation in the higher logarithm approximations. The analyse of the function, which can enter in the expression for such result given in our paper may be used for the reconstruction of the full answer in the next-to-next-to-next-to-leading logarithm approximation (NNNLLA).

Note, that we hope to finish the seven loop computations of the anomalous dimension of the twist-2 operators in the planar $\cN=4$ SYM theory very soon and this result will provide us with very strong test of our result~(\ref{BFKLNNLLA}).

However, it is more interesting to obtain the similar result for QCD. The result, obtained in $\cN=4$ SYM theory, is the most complicated part for the corresponding result in QCD, as can be seen from the comparison of Eq.~(\ref{l12a}) and Eq.~(\ref{BFKLNLLA}). Using the method, presented in this paper, and the available information we are working now under the reconstruction of the eigenvalue of the kernel of BFKL equation in the NNLLA in QCD.

\section*{Acknowledgements}

I would like to thank L.N. Lipatov and V.S. Fadin for useful discussions and for the collaboration in the earlier stage of this work and C. Marboe and D. Volin for the collaboration in the computation of six-loop anomalous dimension without which this work could not be finished.
This research is supported by a Marie Curie International Incoming Fellowship within the 7th European Community Framework Programme, grant number PIIF-GA-2012-331484, by DFG SFB 647 ``Raum -- Zeit -- Materie. Analytische und Geometrische Strukturen'', by RSF grant 14-22-00281 and by RFBR grant 13-02-01246-a.

\section*{Appendix}

In this Appendix we give the list of the functions, which, as we believe, form the basis for the eigenvalue of the kernel of BFKL equation in the next-to-next-to-leading logarithm approximation. We provide their definition and small $\gamma$ expansion up to necessary order.

We suggest, that in general such functions should have the following form
\begin{equation}
\FLS_{-1,a,b,c,\cdots}(\gamma)=
\sum_{k=0}^{\infty }\left(\frac{(-1)^{k+1}}{k-\gamma/2}
+\frac{(-1)^{k+1}}{k+1+\gamma/2}\right)
\Psi_{a,b,c,\cdots}(k+1)\,,
\end{equation}
where $\Psi_{a,b,c,\cdots}(k+1)$ is a generalised $\Psi$ function~\cite{Kotikov:2005gr}, defined as:
\begin{equation}
\Psi_{a,b,c,\cdots}(n+1)=
\sum_{l=0}^{\infty }
\frac{{\mathrm{sign}}(a)^{l+1}}{(l+n+1)^a}
\overline S_{b,c,\cdots}^\pm(l+n+1)\,,
\end{equation}
and $\overline S_{b,c,\cdots}^\pm(l+n+1)$ is the analytical continued harmonic sum (see details in Ref.~\cite{Kotikov:2005gr})
The analytical continuation of $\overline S_{b,c,\cdots}^\pm(l+n+1)$ contains 
$\Psi_{c,\cdots}(k+1)$ and so on. To make the definition of our functions simpler we used the relations between the harmonic sums and arrange the indices in the obtained sums in a such way, that the first indices are negative, while all other are positive. If only the first index is negative, the $\Psi_{a,b,c,\cdots}(k+1)$ will looks like the generalisation of the $\beta(z)$~(\ref{beta0}) or $\beta'(z)$~(\ref{beta}), that is they will have the general future of Eq.~(\ref{HSbeta}) - they will have the product of two $(-1)^k$ factors. For example, to construct $\FLS_{-1,1,-2,1}(\gamma)$, which is the analogy of $S_{1,1,-2,1}$, we use the relation
\begin{equation}
S_{1,-2,1}=-2S_{-2,1,1}+S_1S_{-2,1}+S_{-3,1}+S_{-2,2}
\end{equation}
and define this function in the following way using the second term $S_1S_{-2,1}$ in the above equation:
\begin{eqnarray}
\FLS_{-1,1,-2,1}(\gamma)&=&
\sum_{k=0}^{\infty }\left(\frac{(-1)^{k}}{k-\gamma/2}
+\frac{(-1)^{k}}{k+1+\gamma/2}\right)
\Big(\Psi(k+1)-\Psi(1)\Big)
\times\nonumber\\&&
\qquad\times
(-1)^{k}
\sum_{n=0}\frac{(-1)^{n+1}}{(n+k+1)^2}\Big(\Psi(n+1+k+1)-\Psi(1)\Big)\nonumber\\
&&\hspace*{-25mm}=\
\frac{13}{2} {\z5}
-10 {\z2} {\z3}
+\gamma  \left(
\frac{307}{24} {\z6}
-\frac{41}{4}{\z3}^2
\right)
+\gamma ^2 \left(
-\frac{67}{4} {\z2} {\z5}
+11 {\z3} {\z4}
+\frac{823}{64} {\z7}
\right).
\end{eqnarray}
Define $\FLS_{-1,-2,1,1}(\gamma)$ we use $2S_{1,1}=S_1^2+S_2$
\begin{eqnarray}
\FLS_{-1,-2,1,1}(\gamma)&=&
\sum_{k=0}^{\infty }\left(\frac{(-1)^{k}}{k-\gamma/2}
+\frac{(-1)^{k}}{k+1+\gamma/2}\right)
(-1)^{k}\beta'(k+1)
\Big(\Psi(k+1)-\Psi(1)\Big)^2
\nonumber\\
&=&
\frac{1}{\gamma }\Big(11 {\z4}-16\, {\h31}\Big)
+8 {\z2} {\z3}
-28 {\z5}
+\gamma  
\left(
8\, {\h31} {\z2}
-\frac{179 {\z3}^2}{8}
+\frac{769 {\z6}}{24}
\right)\nonumber\\
&&\quad
+\gamma ^2 \Big(
2 {\z2} {\z5}
+5 {\z3} {\z4}
-12 {\z7}\Big)
\end{eqnarray}
For definition $\FLS_{-1,2,-2}(\gamma)$ we use $S_{2,-2}=S_2S_{-2}+S_{-4}-S_{-2,2}$ and take the first term
\begin{eqnarray}
\FLS_{-1,2,-2}(\gamma)&=&
\sum_{k=0}^{\infty }\left(\frac{(-1)^{k}}{k-\gamma/2}
+\frac{(-1)^{k}}{k+1+\gamma/2}\right)
(-1)^{k}\beta'(k+1)
\Big(\Psi'(k+1)-\Psi'(1)\Big)
\nonumber\\
&=&
6 {\z2} {\z3}
-5 {\z5}
+\gamma  
\Big(
-48\, \h31 {\z2}
+\frac{63}{2} {\z3}^2
-\frac{147}{4} {\z6}
\Big)\nonumber\\&&\quad
+\gamma ^2 
\Big(
-\frac{5}{8} {\z2} {\z5}
+2 {\z3} {\z4}
-\frac{21}{32}{\z7}
\Big)
\end{eqnarray}
and for $\FLS_{-1,1,-3}(\gamma)$ we use $S_{1,-3}=S_1S_{-3}+S_{-4}-S_{-3,1}$ and again take the first term
\begin{eqnarray}
\FLS_{-1,1,-3}(\gamma)&=&
\sum_{k=0}^{\infty }\left(
\frac{(-1)^{k}}{k-\gamma/2}
+\frac{(-1)^{k}}{k+1+\gamma/2}\right)
(-1)^{k}\beta''(k+1)
\Big(\Psi(k+1)-\Psi(1)\Big)
\nonumber\\
&=&
\frac{35}{2} {\z5}
-10 {\z2} {\z3}
+\gamma  
\Big(
\frac{427}{24} {\z6}
-\frac{51}{4} {\z3}^2
\Big)
+\gamma ^2 
\Big(
\frac{1}{2}{\z3} {\z4}
-\frac{67}{4} {\z2} {\z5}
+\frac{1757 }{64}{\z7}
\Big)\,.
\end{eqnarray}
The following sums are defined directly
\begin{eqnarray}
\FLS_{-1,-2,2}(\gamma)&=&
\sum_{k=0}^{\infty }\left(\frac{(-1)^{k}}{k-\gamma/2}
+\frac{(-1)^{k}}{k+1+\gamma/2}\right)
\times\nonumber\\&&
\qquad\times
(-1)^{k}
\sum_{n=0}\frac{(-1)^{n+1}}{(n+k+1)^2}\Big(\Psi'(n+k+2)-\Psi'(1)\Big)\nonumber\\
&=&
\frac{2}{\gamma }\Big(11 {\z4}-32\, \h31\Big)
+\frac{13}{2} {\z5}
-10 {\z2} {\z3}
+\gamma  
\Big(
32\, \h31 {\z2}
-\frac{51}{2} {\z3}^2
+\frac{107 }{3}{\z6}
\Big)\nonumber\\&&\quad
+\gamma ^2 
\Big(
16 {\z2} {\z5}
+8 {\z3} {\z4}
-40 {\z7}
\Big)
\\
\FLS_{-1,-3,1}(\gamma)&=&
\sum_{k=0}^{\infty }\left(\frac{(-1)^{k}}{k-\gamma/2}
+\frac{(-1)^{k}}{k+1+\gamma/2}\right)
\times\nonumber\\&&
\qquad\times
(-1)^{k}
\sum_{n=0}\frac{(-1)^{n+1}}{(n+k+1)^3}\Big(\Psi(n+k+2)-\Psi(1)\Big)\nonumber\\
&=&
\frac{4}{\gamma }\Big(7 {\z4}-8 {\h31}\Big)
+16 {\z2} {\z3}
-48 {\z5}
+\gamma  
\Big(
16 {\h31} {\z2}
-\frac{195}{4} {\z3}^2
+\frac{437 }{6}{\z6}
\Big)\nonumber\\&&\quad
+\gamma ^2 
\Big(
4 {\z2} {\z5}
+4 {\z3} {\z4}
-16 {\z7}
\Big)
\\
\FLS_{-1,-4}(\gamma)&=&
\sum_{k=0}^{\infty }\left(
\frac{(-1)^{k}}{k-\gamma/2}
+\frac{(-1)^{k}}{k+1+\gamma/2}\right)
\frac{(-1)^{k}}{3!}\beta'''(k+1)
\nonumber\\
&=&
\frac{28}{\gamma }\, {\z4}
-16 {\z5}
+\gamma  
\Big(
\frac{145}{6} {\z6}
-12 {\z3}^2
\Big)
-4 {\z7} \gamma ^2
\end{eqnarray}

For the functions with the transcedentality less then $5$ we should expand up to $\gamma^3$ and make such expansion separately for the expansion generated by arguments $(-\gamma/2)$ and $(1+\gamma/2)$ as this function will be multiplied by other functions:
\begin{eqnarray}
\FLS_{-1,-2,1}(\gamma)&=&
\sum_{k=0}^{\infty }\left(
\mathfrak{h}\;\frac{(-1)^{k}}{k-\gamma/2}
+\mfah\;\frac{(-1)^{k}}{k+1+\gamma/2}\right)
\times\nonumber\\&&
\qquad\times
(-1)^{k}
\sum_{n=0}\frac{(-1)^{n+1}}{(n+k+1)^2}\Big(\Psi(n+k+2)-\Psi(1)\Big)\nonumber\\
&=&
\mathfrak{h} 
\bigg[
\frac{20 }{\gamma } {\z3}
-36 {\z4}
-24\, \h31
+42 \ln\! 2\, {\z3}
+\gamma  
\Big(
\frac{29}{4} {\z5}
-5 {\z2} {\z3}
\Big)\nonumber\\&&\quad
+\gamma ^2 
\Big(
\frac{139}{16} {\z3}^2
-\frac{643}{48} {\z6}
\Big)
+\gamma ^3 
\Big(
-\frac{67 }{8}{\z2} {\z5}
+\frac{5 }{2}{\z3} {\z4}
+\frac{1335 }{128}{\z7}
\Big)
\bigg]\nonumber\\&&
+\mfah
\bigg[
16 {\z4}
+24\, \h31
-42 \ln\!2\, {\z3}
+\gamma  
\Big(
\frac{125}{4} {\z5}
-13 {\z2} {\z3}
\Big)\nonumber\\&&\quad
+\gamma ^2 
\Big(
\frac{307}{48} {\z6}
-\frac{107}{16} {\z3}^2
\Big)
+\gamma ^3 
\Big(
-\frac{83}{8} {\z2} {\z5}
+\frac{1}{2}{\z3}{\z4}
+\frac{2359}{128} {\z7}
\Big)
\bigg]
\\
\FLS_{-1,-3}(\gamma)&=&
\sum_{k=0}^{\infty }\left(
\mathfrak{h}\frac{(-1)^{k}}{k-\gamma/2}
+\mfah\frac{(-1)^{k}}{k+1+\gamma/2}\right)
\frac{(-1)^{k}}{(-2)}\;\beta''(k+1)
\nonumber\\
&=&
\mathfrak{h}
\bigg[
\frac{24}{\gamma }\, {\z3}
-16\, \h31
-22 {\z4}
+28 \ln\!2\, {\z3}
+\gamma
\Big(
\frac{67}{2} {\z5}
-18 {\z2} {\z3}
\Big)\nonumber\\&&\quad
+\gamma ^2 
\Big(
\frac{9}{8} {\z3}^2
-2 {\z6}
\Big)
+\gamma ^3 
\Big(
-\frac{75}{4} {\z2} {\z5}
-\frac{3}{2} {\z3} {\z4}
+\frac{2141}{64} {\z7}
\Big)
\bigg]\nonumber\\&&
+\mfah
\bigg[
16\, \h31
+6 {\z4}
-28 \ln\!2\, {\z3}
+\gamma  
\Big(
\frac{83}{2} {\z5}
-18 {\z2} {\z3}
\Big)\nonumber\\&&\quad
+\gamma ^2 
\Big(
-\frac{9}{8} {\z3}^2
-2 {\z6}
\Big)
+\gamma ^3 
\Big(
-\frac{75}{4} {\z2} {\z5}
-\frac{3}{2} {\z3} {\z4}
+\frac{2269}{64} {\z7}
\Big)
\bigg]
\\
\FLS_{-1,1,-2}(\gamma)&=&
\sum_{k=0}^{\infty }\left(
\mathfrak{h}\frac{(-1)^{k}}{k-\gamma/2}
+\mfah\frac{(-1)^{k}}{k+1+\gamma/2}\right)
(-1)^{k}\beta'(k+1)
\Big(\Psi(k+1)-\Psi(1)\Big)
\nonumber\\
&=&
\mathfrak{h}
\bigg[
-26 {\z4}
-24\, \h31
-12 (\ln\!2)^2\; {\z2}
+42 \ln\!2\; {\z3}
+\gamma  
\Big(
\frac{31}{2} {\z2} {\z3}
-\frac{125}{4} {\z5}
\Big)\nonumber\\&&\
+\gamma ^2 
\Big(
-6\, \h31\, {\z2}
+\frac{211}{16} {\z3}^2
-\frac{895}{48} {\z6}
\Big)
+\gamma ^3 
\Big(
\frac{241}{32} {\z2} {\z5}
+\frac{33}{8} {\z3} {\z4}
-\frac{2359}{128} {\z7}
\Big)
\bigg]\nonumber\\&&
+\mfah
\bigg[
+19 {\z4}
+24\, \h31
+12 (\ln\!2)^2\; {\z2}
-42 \ln\!2\; {\z3}
+\gamma  
\Big(
\frac{105}{4} {\z5}
-\frac{27}{2} {\z2} {\z3}
\Big)\nonumber\\&&\
+\gamma ^2 
\Big(
6\, \h31\, {\z2}
-\frac{131}{16} {\z3}^2
+\frac{32}{3}{\z6}
\Big)
+\gamma ^3 
\Big(
-\frac{153}{32} {\z2} {\z5}
-\frac{7}{8} {\z3} {\z4}
+\frac{1183}{128} {\z7}
\Big)
\bigg]
\end{eqnarray}
where for the last function we use $S_{1,-2}=S_1S_{-2}+S_{-3}-S_{-2,1}$.
For the function with the transcedentality $3$ we need the expansion up to $\gamma^4$:
\begin{eqnarray}
\FLS_{-1,-2}(\gamma)&=&
\sum_{k=0}^{\infty }\left(
\mathfrak{h}\frac{(-1)^{k}}{k-\gamma/2}
+\mfah\frac{(-1)^{k}}{k+1+\gamma/2}\right)
(-1)^{k}\beta'(k+1)
\nonumber\\
&=&
\mathfrak{h}
\bigg[
\frac{16}{\gamma } {\z2}
-26 {\z3}
+24 \ln\!2\, {\z2}
-\frac{3}{2} {\z4} \gamma 
+\gamma ^2
\Big(
\frac{21}{2} {\z2} {\z3}
-\frac{83}{4} {\z5}
\Big)\nonumber\\&&\quad
+\gamma ^3 
\Big(
\frac{3}{2} {\z3}^2
-\frac{239}{96} {\z6}
\Big)
+\gamma ^4 
\Big(
\frac{165}{32} {\z2} {\z5}
+\frac{3}{2} {\z3} {\z4}
-\frac{1387}{128} {\z7}
\Big)
\bigg]\nonumber\\&&
+\mfah
\bigg[
10 {\z3}
-24 \ln\!2\, {\z2}
+\frac{13}{2} {\z4} \gamma 
+\gamma ^2 
\Big(
\frac{67}{4} {\z5}
-\frac{21}{2} {\z2} {\z3}
\Big)\nonumber\\&&\quad
+\gamma ^3 
\Big(
\frac{3 }{2}{\z3}^2
-\frac{47 }{96}{\z6}
\Big)
+\gamma ^4 
\Big(
-\frac{165 }{32}{\z2} {\z5}
-\frac{3 }{2}{\z3}{\z4}
+\frac{1259}{128} {\z7}
\Big)
\bigg]
\end{eqnarray}

However, during such simplification some terms can appear as we use the relations for the harmonic sums, while our functions contains $\Psi$-functions, i.e. the analytically continued harmonic sums. For example, using in such relations $S_{-2}(M)$ instead its analytical continuation: 
\begin{equation}
S_{-2}(M)= S_{-2}(\infty)-(-1)^M\beta'(M+1),
\end{equation}
we can lost $S_{-2}(\infty)=-\z2/2$ and so on. We analyse the possible missing terms and have found that we need the following functions:
\begin{eqnarray}
\z2\,\FLS_{-1,2}(\gamma)&\simeq&
\sum_{k=0}^{\infty }\left(
\frac{(-1)^{k}}{k-\gamma/2}
+\frac{(-1)^{k}}{k+1+\gamma/2}\right)
(-1)^{k}\Big((-1)^k\zeta_{-2}\Big)\Big(\Psi'(k+1)-\Psi(1)\Big)
\nonumber\\
&\simeq&
\z2
\bigg(
6 {\z3}
+\gamma  
\Big(
16\, \h31
+{\z4}
\Big)
+\frac{15}{8} {\z5} \gamma ^2
\bigg)\,,
\\
\z3\,\FLS_{-1,-1}(\gamma)&\simeq&
\sum_{k=0}^{\infty }\left(\frac{(-1)^{k}}{k-\gamma/2}
+\frac{(-1)^{k}}{k+1+\gamma/2}\right)
\sum_{n=0}\frac{1}{(n+k+1)}\Big((-1)^{n+k+1}S_{-2,1}(\infty)\Big)\nonumber\\
\nonumber\\
&\simeq&
\z3
\bigg(
\frac{32}{\gamma }\, \ln\!2
-16 {\z2}
+\gamma  
\Big(
18 {\z3}
-16 \ln\!2\, {\z2}
\Big)
-4 {\z4} \gamma ^2
\bigg)\,,
\end{eqnarray}
Moreover, we should take into consideration the product of the listed functions and the functions with one index
\begin{eqnarray}
\FLS_{1}(\gamma)&=&
\mathfrak{h}
\bigg[
\Psi\left(-\frac{\gamma}{2}\right)
-
\Psi\left(1\right)
\bigg]
+
\mfah
\bigg[
\Psi\left(1+\frac{\gamma}{2}\right)
-
\Psi\left(1\right)
\bigg]
\\
\FLS_{-a}(\gamma)&=&
\sum_{k=0}^{\infty }\left(
\mathfrak{h}
\frac{(-1)^{k+1}}{\big(k-\gamma/2\big)^a}
+\mfah\frac{(-1)^{k+1}}{\big(k+1+\gamma/2\big)^a}\right)
\\
\FLS_{a}(\gamma)&=&
\sum_{k=0}^{\infty }\left(
\mathfrak{h}
\frac{1}{\big(k-\gamma/2\big)^a}
+\mfah\frac{1}{\big(k+1+\gamma/2\big)^a}\right)
\end{eqnarray}
which have the following expansion
\begin{eqnarray}
\FLS_{-1}(\gamma)&=&
\mathfrak{h}
\bigg[
-\frac{2}{\gamma }
-\text{ln2}
-\frac{{\z2} \gamma }{4}
-\frac{3 {\z3} \gamma ^2}{16}
\bigg]
+\mfah
\bigg[
\text{ln2}
-\frac{{\z2} \gamma }{4}
+\frac{3 {\z3} \gamma ^2}{16}
\bigg]\nonumber\\
\FLS_{1}(\gamma)&=&
\mathfrak{h}
\bigg[
-\frac{{\z2} \gamma }{2}
-\frac{{\z3} \gamma ^2}{4}
+\frac{2}{\gamma }
\bigg]
+\mfah 
\bigg[
\frac{{\z2} \gamma }{2}
-\frac{{\z3} \gamma ^2}{4}
\bigg]
\nonumber\\
\FLS_{-2}(\gamma)&=&
\mathfrak{h}
\bigg[
-\frac{{\z2}}{2}
-\frac{3 {\z3} \gamma }{4}
-\frac{21 {\z4} \gamma ^2}{32}
+\frac{4}{\gamma ^2}
\bigg]
+
\mfah
\bigg[
\frac{{\z2}}{2}
-\frac{3 {\z3}\gamma }{4}
+\frac{21 {\z4} \gamma ^2}{32}
\bigg]
\nonumber\\
\FLS_{2}(\gamma)&=&
\mathfrak{h}
\bigg[
-{\z2}
-{\z3} \gamma
-\frac{3 {\z4} \gamma ^2}{4}
-\frac{4}{\gamma ^2}
\bigg]+
\mfah
\bigg[
-{\z2}
+{\z3} \gamma
-\frac{3 {\z4} \gamma^2}{4}
\bigg]
\nonumber\\
\FLS_{-3}(\gamma)&=&
\mathfrak{h}
\bigg[
-\frac{3 {\z3}}{4}
-\frac{21 {\z4} \gamma }{16}
-\frac{45 {\z5} \gamma ^2}{32}
-\frac{8}{\gamma ^3}
\bigg]
+
\mfah
\bigg[
\frac{3 {\z3}}{4}
-\frac{21{\z4} \gamma }{16}
+\frac{45 {\z5} \gamma ^2}{32}
\bigg]
\nonumber\\
\FLS_{3}(\gamma)&=&
\mathfrak{h}
\bigg[
-{\z3}
-\frac{3 {\z4} \gamma }{2}
-\frac{3 {\z5} \gamma ^2}{2}
+\frac{8}{\gamma ^3}
\bigg]
+
\mfah 
\bigg[
-{\z3}
+\frac{3 {\z4} \gamma }{2}
-\frac{3{\z5} \gamma ^2}{2}
\bigg]
\nonumber\\
\FLS_{-4}(\gamma)&=&
\mathfrak{h}
\bigg[
-\frac{7 {\z4}}{8}
-\frac{15 {\z5} \gamma }{8}
-\frac{155 {\z6} \gamma ^2}{64}
+\frac{16}{\gamma ^4}
\bigg]
+
\mfah 
\bigg[
\frac{7 {\z4}}{8}
-\frac{15 {\z5} \gamma }{8}
+\frac{155 {\z6} \gamma ^2}{64}
\bigg]
\nonumber\\
\FLS_{4}(\gamma)&=&
\mathfrak{h}
\bigg[
-{\z4}
-2 {\z5} \gamma 
-\frac{5 {\z6} \gamma ^2}{2}
-\frac{16}{\gamma ^4}
\bigg]
+
\mfah
\bigg[
-{\z4}
+2 {\z5} \gamma 
-\frac{5 {\z6} \gamma^2}{2}
\bigg]
\nonumber\\
\FLS_{-5}(\gamma)&=&
\mathfrak{h}
\bigg[
-\frac{15 {\z5}}{16}
-\frac{155 {\z6} \gamma }{64}
-\frac{945 {\z7} \gamma ^2}{256}
-\frac{32}{\gamma ^5}
\bigg]
+
\mfah 
\bigg[
\frac{15 {\z5}}{16}
-\frac{155 {\z6} \gamma }{64}
+\frac{945 {\z7} \gamma ^2}{256}
\bigg]
\nonumber\\
\FLS_{5}(\gamma)&=&
\mathfrak{h}
\bigg[
-{\z5}
-\frac{5 {\z6} \gamma }{2}
-\frac{15 {\z7} \gamma ^2}{4}
+\frac{32}{\gamma ^5}
\bigg]
+
\mfah 
\bigg[
-{\z5}
+\frac{5 {\z6} \gamma }{2}
-\frac{15 {\z7} \gamma ^2}{4}
\bigg]\,.
\end{eqnarray}
Thus, the most common basis contains the following functions:
\begin{eqnarray}
{\mathrm{Basis}}\,\Big|_{\mathrm{NNLLA}}&=&
\Big\{\FLS_{-1,-2,1,1},
\FLS_{-1,1,-2,1},
\FLS_{-1,1,1,-2},
\FLS_{-1,-2,2},
\FLS_{-1,2,-2},
\FLS_{-1,-3,1},
\FLS_{-1,1,-3},
\FLS_{-1,-4},
\nonumber\\[1mm]&&
\quad
{\FLS_{1}\otimes\FLS_{-1,-2,1}},
{\FLS_{1}\otimes\FLS_{-1,1,-2}},
{\FLS_{1}\otimes\FLS_{-1,-3}},
{\FLS_{2}\otimes\FLS_{-1,-2}},
{\FLS_{-2}\otimes\FLS_{-1,-2}},
\nonumber\\[2mm]&&
\quad
{\FLS_{1}\otimes\FLS_{1}\otimes\FLS_{-1,-2}},
{\FLS_{1}\otimes\FLS_{2}\otimes\FLS_{2}},
{\FLS_{1}\otimes\FLS_{-2}\otimes\FLS_{2}},
{\FLS_{1}\otimes\FLS_{-2}\otimes\FLS_{-2}},
\nonumber\\[2mm]&&
\quad
{\FLS_{1}\otimes\FLS_{1}\otimes\FLS_{-3}},
{\FLS_{1}\otimes\FLS_{1}\otimes\FLS_{3}},
{\FLS_{1}\otimes\FLS_{4}},
{\FLS_{1}\otimes\FLS_{-4}},
{\FLS_{2}\otimes\FLS_{3}},
\nonumber\\[2mm]&&
\quad
{\FLS_{-2}\otimes\FLS_{3}},
{\FLS_{2}\otimes\FLS_{-3}},
{\FLS_{-2}\otimes\FLS_{-3}},
{\FLS_{5}},
{\FLS_{-5}},
{\z2\times\FLS_{-3}},
{\z2\times\FLS_{3}},
\nonumber\\[2mm]&&
\quad
{\z2\times\FLS_{-1}\otimes\FLS_{2}},
{\z2\times\FLS_{-1}\otimes\FLS_{-2}},
{\z2\times\FLS_{1}\otimes\FLS_{2}},
{\z2\times\FLS_{1}\otimes\FLS_{-2}},
\nonumber\\[2mm]&&
\quad
{\z2\times\FLS_{-1,-2}},
{\z2\times\FLS_{-1,2}},
{\z3\times\FLS_{2}},
{\z3\times\FLS_{-2}},
{\z3\times\FLS_{-1}\otimes\FLS_{-1}},
{\z3\times\FLS_{-1,-1}},
\nonumber\\[2mm]&&
\quad
{\z4\times\FLS_{1}},
{\z4\times\FLS_{-1}},
\z2\z3,
\z5
\Big\}
\end{eqnarray}
where the operation $\otimes$ means the product of function separately for the holomorphic and antiholomorphic parts, that is, for example,
\begin{equation}
{\FLS_{2}\otimes\FLS_{-3}}=
{\FLS_{2}\!\!\left(-\frac{\gamma}{2}\right)\FLS_{-3}\!\!\left(-\frac{\gamma}{2}\right)}+
{\FLS_{2}\!\!\left(1+\frac{\gamma}{2}\right)\FLS_{-3}\!\!\left(1+\frac{\gamma}{2}\right)}\,.
\end{equation}

\end{document}